\documentclass[sigconf, screen]{acmart}
\acmConference[EASE 2026]{The 30th International Conference on Evaluation and Assessment in Software Engineering}{9–12 June, 2026}{Glasgow, Scotland, United Kingdom}
\AtBeginDocument{%
  }

\setcopyright{acmlicensed}
\copyrightyear{2026}
\acmYear{2026}
\acmDOI{XXXXXXX.XXXXXXX}
\usepackage{textcomp}
\usepackage{xcolor}
\usepackage{xspace}
\usepackage[disable]{todonotes}
\usepackage{url}
\usepackage{hyperref}
\usepackage{array}
\usepackage{enumitem}
\definecolor{myPurple}{RGB}{128, 0, 128} 
\definecolor{myBlue}{RGB}{30, 144, 255}   
\definecolor{myGreen}{RGB}{200,0,0}   
\definecolor{myRed}{RGB}{220, 20, 60} 
\definecolor{rq1Color}{RGB}{220, 20, 60} 
\begin{document}

\title{Integrating Log-Based Security Analytics in Agile Workflows:\\ A Real-World Experience Report}

\author{Arpit Thool}
\email{arpitthool@vt.edu}
\orcid{0009-0003-8132-2887}
\affiliation{%
  \institution{Virginia Tech}
  \city{Blacksburg}
  \state{Virginia}
  \country{USA}
}

\author{Chris Brown}
\email{dcbrown@vt.edu}
\orcid{0000-0002-6036-4733}
\affiliation{%
  \institution{Virginia Tech}
  \city{Blacksburg}
  \state{Virginia}
  \country{USA}
}


\begin{abstract}
Modern organizations increasingly rely on log data and monitoring signals to protect products against account takeovers and abuse, yet integrating security analytics into fast-moving Agile workflows remains challenging. While it is important to understand how security practices are developed and sustained within Agile, real-world case studies of such integrations remain scarce. 
This experience report provides insights on developer perceptions of an effort to integrate log-based fraud detection within an organization, known as the ``Red Flag Project''. A cross-functional team of eight members (including one author) iterated weekly to implement a proof-of-concept log-based system that alerts stakeholders when accounts exhibit suspicious activity patterns. Through semi-structured interviews, we investigate developer perceptions of log-based fraud detection integration---exploring their willingness to adopt the system, challenges encountered, and the overall impact on day-to-day development activities and security perceptions. Our findings highlight key lessons, mitigation techniques, and best practices for embedding security analytics into Agile workflows. We provide insights for practitioners and researchers seeking to incorporate security practices into modern development processes while maintaining both speed and resilience in software delivery.

\end{abstract}

\ccsdesc[500]{Security and privacy~Web application security}
\ccsdesc[500]{Security and privacy~Software security engineering}
\ccsdesc[500]{Security and privacy~Social aspects of security and privacy}
\ccsdesc[500]{Security and privacy~Usability in security and privacy}
\ccsdesc[500]{Security and privacy~Vulnerability scanners}
\ccsdesc[300]{Software and its engineering~Software maintenance tools}
\ccsdesc[300]{Software and its engineering~Software verification and validation}


\keywords{Agile, Security, Use Case Study, Fraud Detection, Splunk, Developer Perspectives}


\maketitle

\section{Introduction}

Modern software products are increasingly deployed as \textit{web applications}~\cite{jazayeri2007some,andreessen2011software}, and developed using iterative development approaches such as Agile and continuous integration/continuous delivery (CI/CD)~\cite{hoda2018rise,techreport,rahman2015synthesizing}.
While these practices enable rapid iteration and frequent releases, prior work shows that integrating security practices into these workflows can be difficult due to process misalignment, tool friction, and documentation overhead~\cite{rindell2018aligning,rajapakse2021empirical,bajpai2022secure}. This tension is well documented across automated security tools, including Dynamic Application Security Testing (DAST)~\cite{arpit-use-case} and Static Application Security Testing (SAST)~\cite{bolduc2016lessons} systems.

Organizations also rely on logs to safeguard user accounts against abuse, fraud, and account takeovers (ATOs)~\cite{verizon,nsa}. Log-based security analytics---e.g. mining authentication events, session anomalies, and behavioral signals---offer a complementary and runtime-oriented approach to defense. For instance, Splunk\footnote{\url{https://www.splunk.com/}} is a widely used log analytics and Security Information and Event Management (SIEM) platform for ingesting, indexing, querying, and visualizing log data at scale~\cite{subramanian2020introducing}. Unlike code and program analysis techniques, log-driven detection leverages production events to surface suspicious activity patterns and trigger timely alerts~\cite{nour2023survey}. However, the integration of such activities and day-to-day impact of log-based security analytics into Agile workflows remains unexplored within industrial contexts.

This paper reports an experience report conducted within the Information Technology (IT) division of an anonymized organization that experienced a real-world security incident involving unauthorized access to accounts. To mitigate the risk of further account takeovers and abuse, the leadership initiated the ``Red Flag Project''---a proof-of-concept effort to integrate the security practice of log-based fraud detection through the design and implementation of a solution that involved Splunk, a Python (Flask) middleware and Grouper---an Identity Access Management (IAM) tool. The project aimed to detect suspicious activity patterns in production authentication logs and promptly alert stakeholders such as the billing team, enabling a faster response to malicious behavior.

The first author served as an embedded software engineer in this team. After the rollout, we conducted semi-structured interviews with seven team members to understand their willingness to adopt, perceived impacts, and opportunities for process improvement. Our work complements experience reports on integrating security into Agile pipelines (e.g., DAST in Kanban/CI/CD)~\cite{arpit-use-case,arpit-dast} and broadens evidence on how practitioners perceive security activities in Agile environments~\cite{thool2026bridging}.

\noindent We investigate the following research questions:
\begin{itemize}[topsep=2pt,itemsep=1pt,leftmargin=12pt]
    \item[] \textbf{RQ1} \label{rq:adoption} How do team members perceive the integration of log-based fraud detection capabilities into their Agile (Kanban) workflow, and what factors influence their acceptance and continued use of this security-oriented analytics practice?
    \item[] \textbf{RQ2} \label{rq:challenges} What challenges emerge when integrating a log-based fraud detection system within an Agile development environment, and how do these challenges affect team productivity, security awareness, and the overall development process?
    \item[] \textbf{RQ3 } \label{rq:improvement} What strategic modifications to Agile workflows and team practices can enhance the integration of log-based fraud detection systems?
\end{itemize}

Our findings revealed all team members were consistently willing to adopt and continue using the log-based fraud detection system, highlighting its usefulness in correlating signals and enhancing awareness. Day-to-day impact was minimal, though challenges included coordination overhead, system fragility, and reliance on a dedicated engineer. Our findings make three key contributions: \emph{(i)} A practice-grounded experience report on designing, delivering, and operating a \emph{log-based fraud detection} capability within an Agile (Kanban) team, from initial incident response to production deployment. \emph{(ii)} An empirical account of practitioner perceptions regarding \emph{adoption, friction points, and perceived impacts} of integrating security analytics into fast-moving workflows, complementing prior studies on security-in-Agile and tool integration~\cite{rindell2018aligning,rajapakse2021empirical,bajpai2022secure,arpit,arpit-use-case}. \emph{(iii)} \todo{refine more} Actionable \emph{process recommendations} (e.g., backlog triage patterns, alert design, roles/rituals, and feedback loops) to sustain continued use of log-driven security analytics without undermining team velocity or developer experience.


\section{Background}
\label{section-background}

\paragraph{\textit{\textbf{Agile:}}}
\label{section-background-agile}
 Agile software development is the predominant approach for organizing modern software teams, emphasizing iterative delivery, responsiveness to change, and close collaboration~\cite{hoda2018rise,techreport}. Among Agile methods, \textit{Kanban} is widely adopted for its continuous flow, explicit work-in-progress (WIP) limits, and visual workflow management that reduces bottlenecks and supports frequent releases~\cite{hofmann2018development,winska2020software}. While these properties help teams sustain delivery velocity, prior work shows integrating security activities into Agile pipelines can be difficult due to misaligned cadences, additional documentation burdens, and tool friction~\cite{rindell2018aligning,rajapakse2021empirical,bajpai2022secure}.

\textbf{\emph{Security and Fraud Detection:}}
\label{section-background-security}
Web applications face persistent threats such as account takeover (ATO), abuse, and data exfiltration, keeping application security a critical concern~\cite{alzahrani2017web}. A broad toolbox exists for securing software, spanning \textit{design-time} techniques (e.g., SAST, threat modeling) and \textit{run-time} techniques (e.g., DAST, monitoring)~\cite{dupont2021incremental,arpit-dast}. Whereas static and dynamic testing primarily target vulnerabilities before release, \emph{log-driven} fraud and anomaly detection leverages production telemetry (authentication events, session metadata, request patterns) to surface suspicious behaviors as they occur, enabling faster detection and response. In this work, we use ``fraud detection'' to refer to suspicious or abusive account activity patterns (e.g., credential stuffing indicators, anomalous geo/login sequences) discovered via log analytics in production systems.

\textit{\textbf{Experience Report Setting:}}
\label{sec:setting}
The ``Red Flag Project'' team was composed of 8 members drawn from different functional roles, including engineers, analysts, and managers. The first author participated as a software engineer embedded in this team.
This cross-functional group was intentionally formed to bring together diverse perspectives.  
The team worked in the same timezone and collaborated remotely. Membership was stable throughout the project lifecycle, supporting iterative decision making and continuity.
The team adopted a Kanban-style Agile process, characterized by continuous flow, visual workflow tracking, and work-in-progress (WIP) limits. Weekly meetings served as the primary cadence for prioritizing tasks, progress reviews, backlog refinement, and discussion of new ``Red Flag'' heuristics. During each weekly call, ad hoc discussions were encouraged when specific blockers or design decisions required additional attention.
The solution was implemented iteratively and deployed in production as version 1.0. Development and operations leveraged Git-based version control and CI/CD pipelines to streamline integration and deployment of detection queries and alert mechanisms. This Agile environment created suitable conditions for investigating the integration of log-based fraud detection into development workflows~\cite{bartsch2011practitioners}.



\section{Experience Report}
\label{sec:methodology}


To structure our analysis, we draw on the five iterative steps of the \textit{action research cycle}~\cite{petersen2014action,staron2020action}: \textit{problem diagnosis}, \textit{action planning}, \textit{action taking}, \textit{evaluation}, and \textit{specifying learning}. Previous studies leverage this framework to investigate the integration of security practices in Agile development (e.g.,~\cite{arpit-use-case,lwakatare2021experiences}). 

\emph{\textbf{Problem Diagnosis:}}
During this phase, the problem is identified and described in industrial contexts~\cite{petersen2014action}. Industry reports  highlight that a significant proportion of security incidents involve unauthorized account access and fraudulent use of credentials~\cite{verizon,nsa}. Detecting such incidents requires runtime-oriented approaches, as traditional pre-release testing cannot directly address attacks against production accounts~\cite{dupont2021incremental,arpit-dast}. Further, prior work suggests Agile development teams often encounter difficulties integrating security activities into their workflows, stemming from process misalignment, tool friction, and documentation overhead~\cite{arpit,rindell2018aligning,rajapakse2021empirical}. 

Attackers were able to gain unauthorized access to user accounts and this exposed gaps in the organization’s ability to detect account takeovers and prompted leadership to initiate the ``Red Flag Project''.
The team's primary goal was to surface suspicious login behaviors and account misuse by leveraging production authentication logs via Splunk.  
The team collaboratively defined a set of ``Red Flag'' heuristics (e.g., rapid login failures, unusual geo-location changes, device churn) and prioritized which ones should be implemented first. The first author, embedded as a developer within the team, was responsible for implementing these heuristics in Splunk.

\emph{\textbf{Action Planning:}}
This phase examines potential approaches for addressing the problem~\cite{petersen2014action}. The plan was co-created with the researcher and the team, balancing the organization’s infrastructure, resources, and urgent need to detect fraudulent account activity. The team identified three suspicious behaviors to serve as initial detection heuristics: \textit{(a)} a user failing multi-factor authentication (MFA) more than twice in a 24-hour period; \textit{(b)} a user creating or modifying \hyperref[https://www.microsoft.com/en-gb/microsoft-365/outlook]{Outlook} rules that intercept or redirect direct-deposit–related emails; and \textit{(c)} a user attempting to change their payroll direct-deposit information; as these are the events that were observed to be common with all the attackers.

Two primary challenges emerged during planning. First, the system needed to \emph{aggregate} these signals so that if a user triggered all three indicators, they would be elevated as a high-risk account. Second, stakeholders required a reliable \emph{notification mechanism} to receive timely alerts when such high-risk accounts were identified.  
To address aggregation, the team leveraged Grouper\footnote{\url{https://incommon.org/software/grouper/}}---an open-source access management tool that centralizes the administration of group- and role-based authorization across organizational systems. Grouper supports time-bound group memberships and nested group hierarchies and manages ephemeral memberships (e.g., auto-expiration of MFA failure flags after 24 hours). We aggregate them into a final ``Red Flag’’ group that dynamically reflected high-risk accounts. 
For notifications, we developed a lightweight mechanism to periodically check the final ``Red Flag'' group and inform stakeholders of new entries via email. To implement this, we used a custom Python script with existing email utilities (i.e., Mutt\footnote{\url{http://www.mutt.org/}}) triggered through scheduled jobs.  

As Splunk and Grouper do not directly interface with each other, the team implemented an intermediary web service to bridge these systems. A Flask-based application hosted on an internal Red-Hat server was selected to receive Splunk alerts via webhooks \footnote{\hyperref[https://help.splunk.com/en/splunk-enterprise/alert-and-respond/alerting-manual/9.3/configure-alert-actions/use-a-webhook-alert-action]{Splunk Webhooks}} and programmatically insert flagged accounts into the appropriate Grouper groups. This satisfied the constraints of reusing existing infrastructure, minimizing manual overhead, and enabling end-to-end automation of detection and alerting. 


\emph{\textbf{Action Taking:}}
This phase focuses on executing the planned actions~\cite{petersen2014action}. The researcher carried out the intervention as a \textit{direct action}~\cite{staron2020action}, actively influencing the organization’s processes.
The first author developed and deployed the Flask web service. Splunk alerts corresponding to the three fraud detection heuristics were configured to trigger webhooks to this service. Upon receiving a webhook, the Flask application extracted the relevant account entry and invoked Grouper’s API to insert the entry into the appropriate detection group (e.g., MFA failures, payroll changes, Outlook rules). 
Grouper then maintained these memberships for a bounded duration (e.g., 24 hours), automatically expiring entries after the time window elapsed. A final ``Red Flag’’ group aggregated the three individual detection groups. If an account was present in all three, it was automatically promoted into the Red Flag group. Conversely, if a membership expired the account was removed from the final group, ensuring that only actively suspicious accounts persisted. 
To implement stakeholder notifications, the first author developed a Python script scheduled via a Cron job. With appropriate credentials the script queried the final ``Red Flag'' group for new entries added since its last execution. If new accounts were detected, the script automatically generated and sent an email notification to stakeholders using Mutt. This ensured that alerts were both timely and minimally redundant, surfacing only newly flagged accounts rather than re-reporting previously handled entries.  
Through this implementation, the ``Red Flag'' team delivered version 1.0 of the fraud detection capability into production, consisting of: 
\textcolor{blue}{$\text{Grouper} \rightarrow \text{Python Script (Cron)} \rightarrow \text{Google Group (Email Notification)}$}.


\emph{\textbf{Evaluation:}}
The evaluation phase examines the effects of the captured actions~\cite{petersen2014action}. 
Our study employed focused observation~\cite{stausberg2011structured}, a field technique where the observer concentrates on particular phenomena in their natural environment to obtain detailed insights.
We applied this approach with high researcher involvement and minimal team awareness of observation~\cite{runeson2009guidelines}. The first author actively participated as a team member, attending weekly online Zoom meetings and ad hoc design discussions on Slack. These activities were observed during regular workflows to minimize disruption. Participant confidentiality and anonymity were maintained, and the study was approved by the project manager.
We conducted \textit{qualitative interviews} with team members to reflect on their experiences integrating the log-based fraud detection capability into their Agile workflow. Our interview protocol was reviewed and approved  by the Institutional Review Board (IRB) to ensure participant confidentiality and compliance with organizational policies.


\begin{table}[t] \label{table-participants}
    \small
    \centering
    \caption{Team Members}

    \begin{tabular}{ m{1.5cm} | m{3.5cm} | m{1.4cm} } 
    \toprule
     \textbf{Participant} & \textbf{Role} & \textbf{Industry Exp. (Years)}   \\ 
\midrule

 M1\label{M1} 
& 	Senior Escalation Engineer	& 30
\\ \hline

M2\label{M2}  
& IT Security Analyst & 16
\\ \hline

M3\label{M3} 
& 	Applications Engineer II & 20
\\ \hline

M4\label{M4} 
&	Cloud Security Operations Analyst &	7
\\ \hline

M5\label{M5} 
&	Test Engineering Manager & 45
\\ \hline

M6\label{M6} 
&	Director of Cyber Defense &	30
\\ \hline

M7\label{M7} 
&	Senior IT program manager & 30
\\ \hline

M8\label{M8} 
&	Software Engineer (author) &	4
\\ 
\bottomrule
    \end{tabular}
    
    \label{table-participants}

\end{table}

\emph{(i) Participants:}
Once the initial fraud detection system was deployed in production, we initiated qualitative interviews to capture practitioner perceptions of adoption, challenges, and process impacts. We distributed an open invitation through the team’s communication channel.
Of the seven (excluding the author) eligible team members, all agreed to take part.
As indicated in Table~\ref{table-participants}, participants represented a range of roles across the project team, including application engineer, security analyst, program manager, and engineering manager. This role diversity allowed us to capture perspectives from both technical and non-technical stakeholders, ensuring a comprehensive understanding of integration effort.
The participants brought substantial professional expertise, with industry experience ranging from 7 to 45 years (average: 24.5 years). 


\emph{(ii) Interview Design:}
Interviews were conducted virtually to accommodate remote collaboration practices. Each session lasted between 30–45 minutes.
The interviews were semi-structured, following a protocol of ten open-ended questions~\cite{Thool2026}. 
These questions were developed to correspond with our research objectives, examining different aspects of the team’s experience. Interviews were conducted via Microsoft-Teams or Zoom, according to participants' preferences, and each session was recorded with consent for subsequent transcription and analysis.
To address \hyperref[rq:adoption]{RQ1}, we asked participants about their initial willingness to adopt the Splunk-based fraud detection system and their comfort level with continuing to use it after integration. 
The next set of questions aligned with \hyperref[rq:challenges]{RQ2}, focusing on the challenges and impacts of fraud detection integration. 
%
Finally, the last section provided data to answer \hyperref[rq:improvement]{RQ3} by eliciting suggestions for process modifications, required resources, and design improvements. 

\emph{(iii) Data Analysis:}
The interviews using the transcribe functionality in Microsoft Word.\footnote{\url{https://word.cloud.microsoft/}}
We applied an open coding process, a standard approach in qualitative research~\cite{fitzpatrick2013us}, to examine the interview transcripts~\cite{blair2015reflexive}. We analyzed responses to understand perceptions of log-based fraud detection---identifying, labeling, and organizing data into themes or patterns without using predefined codes. Two researchers independently coded the responses and then reconciled their results to establish the final categories, deriving insights from the participants’ input.
We published an example of our data analysis via Figshare~\cite{Thool2026}.
 After individual analyses, we coded all responses to each question to derive themes. We are not able to share the raw transcripts due to confidentiality considerations.

\emph{\textbf{Specifying Learning:}}
The final phase involves articulating generalized learning grounded in our evaluation~\cite{petersen2014action}. Drawing on the interview outcomes, we distill lessons for integrating \emph{log-based security analytics} into Agile (Kanban) workflows---covering role/ownership patterns, automation, and governance for sustained use.  Section~\ref{sec:discussion} consolidates these implications to guide future teams considering fraud-detection integration in modern development processes.

\section{Results}\label{sec:results}

\subsection{\hyperref[rq:adoption]{RQ1}: Integration Willingness}

\paragraph{\textit{\textbf{Original Willingness:}}}
All interviewees expressed a \textcolor{rq1Color}{Willingness} (6/7) to adopt the Splunk-based fraud detection system into their Agile workflow. They consistently framed the capability as both \emph{useful} and a \emph{welcome change}. 
For example, \hyperref[M7]{M7} noted they were \emph{``very much so willing to implement it into agile''}.  
Despite willingness, participants also raised concerns about \textcolor{rq1Color}{Long-Term Sustainability} (2/7), such as risks of abandonment, unclear ownership, and ongoing funding requirements. As \hyperref[M5]{M5} warned, \emph{``ongoing maintenance and support… funding and resources… keeping stuff up to date''}. Technical fragility was also raised, particularly around Splunk’s reliance on stable log formats.
Others highlighted organizational or process-level risks, such as unclear scope and requirements (\hyperref[M2]{M2}), difficulties aggregating signals from multiple sources (\hyperref[M4]{M4}).

\textit{\textbf{Continued Willingness:}}
After deployment, all participants expressed \textcolor{rq1Color}{comfort} (6/7) with continuing to use the system,
though some emphasized \textcolor{rq1Color}{Conditional Comfort} (3/7) tied to system robustness and ownership. \hyperref[M3]{M3} highlighted the importance of sustained oversight: \emph{``As long as somebody is in charge of it and keeps an eye on it''}. 
\textcolor{rq1Color}{Technical Fragility and Scaling Concerns} (3/7) persisted post-deployment, with participants stressing the need for tuning and maintenance. For example, \hyperref[M1]{M1} cautioned that \emph{``this particular build is a little fragile''}. 
Others included latency in signal-to-alert pipelines (\hyperref[M4]{M4}), fragility of log formats (\hyperref[M5]{M5}), and the need for periodic \emph{keep-alive} testing (\hyperref[M5]{M5}).
\textcolor{rq1Color}{Process Clarity Concerns} (1/7) for downstream stakeholders also emerged, as \hyperref[M2]{M2} questioned whether recipients of notifications would know how to act: \emph{``The detection has been sending emails, but those emails might need to be finessed… have we communicated what we want you to do?''}
On the positive side, some participants framed the \textcolor{rq1Color}{System as a Foundation} (1/7) for broader future capabilities. \hyperref[M6]{M6} emphasized this forward-looking stance: \emph{``We might be able to expand it to do some other use cases… this one was a pretty tight and narrow focus, but it shows we can actually develop this sort of thing''}.  

\subsection{\hyperref[rq:challenges]{RQ2}: Challenges and Impact}

\paragraph{\textit{\textbf{Challenges:}}}
When asked about challenges in incorporating the Splunk-based fraud detection system into their existing workflows, participants reported a mix of organizational, technical, and process-related frictions. The most frequent theme was related to \textcolor{rq1Color}{Project \& Resource Trade-offs} (5/7), with participants noting velocity remained stable only because a dedicated resource was assigned. As \hyperref[M5]{M5} cautioned, once that resource is reassigned, \emph{``we’re going to lose capability''}.
A second theme was \textcolor{rq1Color}{Meeting \& Scheduling Challenges} (5/7). While some characterized the impact as light---\hyperref[M3]{M3} explained it was \emph{``not bad… another short meeting a week was not a problem''}---others noted scheduling conflicts and focus issues. For instance, \hyperref[M7]{M7} cited difficulties in \emph{``keeping the team on time, on schedule, and focused''}, with meetings sometimes derailing into sidebars where \emph{``everybody had an opinion''}. 
Two participants mentioned \textcolor{rq1Color}{Skill Gaps \& Upskilling} (2/7), particularly around Splunk querying and alert creation though these were often reframed as learning opportunities. 
%
Despite these challenges, several participants noted \textcolor{rq1Color}{Positive Impacts} (3/7), emphasizing that automation reduced manual investigation burden and freed the capacity for higher-value work.

\textit{\textbf{Impact on Day-to-Day Development Activities:}}
Across participants, the integration produced \textcolor{rq1Color}{Low/Minimal Day-To-Day Impact} (5/7), with most reporting little to no change in their routine work. 
For example, \hyperref[M2]{M2} noted: \emph{``Minimal impact… attending the meetings was the only real impact''}. 
While \textcolor{rq1Color}{Meeting/Time Costs} (5/7) was acknowledged, they were generally modest.

Two participants emphasized \textcolor{rq1Color}{Cross-Team Coordination \& Latency} (2/7) as the primary source of overhead.
\hyperref[M3]{M3} noted, \emph{``It was sometimes complicated to get all those different teams together''}. \hyperref[M5]{M5} echoed systemic bottlenecks: \emph{``Messaging and decision making is always a lot slower… price of doing business'',} citing \textcolor{rq1Color}{Dependency delays/rollover} \emph{``Waiting on one person to do one thing… push it to next week''}, and the \textcolor{rq1Color}{Need for coordination}. They also recounted \textcolor{rq1Color}{Coordination overhead \& rollback}: \emph{``We went all the way to production… then had to pull back… walk through a few more tests… soon… go back to production''}.
In some cases, \textcolor{rq1Color}{Pre-Existing Detection/Ingestion Pieces} reduced day-to-day change (1/7). \hyperref[M4]{M4} explained that Splunk event detection rules were setup beforehand, so not much changed in their day-to-day: \emph{``We already had sort of bits and pieces lying around… that just took a little bit of development to get working again. Didn’t change day-to-day activities much at all''}.
Finally, one participant anticipated \textcolor{rq1Color}{Future Expansion} potentially increasing involvement (1/7), stating: \emph{``Going forward? I’m now an SME on this service… we’re going to add other alerts… I expect to be involved in both the Splunk side and the… side of moving the data into something that can then make that decision… it’s a thing on my list''} (\hyperref[M1]{M1}). 

\textit{\textbf{Perceived Impact on Security:}}
Participants expressed mixed views on whether the fraud detection system improved overall security. A recurring theme was \textcolor{rq1Color}{Conditional or Limited Improvement} (3/7). For instance, \hyperref[M1]{M1} cautioned, \emph{``No, I do not believe that's happened yet… ''}, emphasizing that effectiveness hinged on producing actionable detection. 
Similarly, \hyperref[M4]{M4} recognized greater awareness but stressed that \emph{``users are still vulnerable as long as people's accounts are getting popped''}. \hyperref[M5]{M5} agreed it was an \emph{``improvement, yes… raises the visibility''}. 
Concerns over \textcolor{rq1Color}{False positives and Accuracy} (2/7) were particularly salient. \hyperref[M2]{M2} described it as their \emph{``main concern''}, especially given the frequency of legitimate direct-deposit changes, but noted that \emph{``in testing… we didn’t detect a high false positive rate''}. They emphasized the trade-off, concluding, \emph{``Even if alerts turn out to be benign, the alternative is someone potentially losing money''}. Likewise, \hyperref[M1]{M1} warned that excess false positives could \emph{``trap overall service because it will take cycles… to do the investigations''}.
At the same time, participants articulated \textcolor{rq1Color}{Clear Security Benefits} (3/7). \hyperref[M3]{M3} stated, \emph{``Yeah, I absolutely think it's improved the overall security''} while \hyperref[M6]{M6} noted: \emph{``Reducing the amount of time it takes to identify when an account has been compromised… that is an improvement''}. Similarly, \hyperref[M7]{M7} was unequivocal: \emph{``Most definitely has affected it in a very positive way… it will make it easier for us to detect fraud''}.

Beyond direct detection, participants highlighted \textcolor{rq1Color}{Awareness and Collaboration Benefits} (3/7). For example, \hyperref[M3]{M3} described how the initiative \emph{``brought together a bunch of teams that hadn’t officially worked together''}, while \hyperref[M4]{M4} noted that simply being notified earlier increased awareness. \hyperref[M5]{M5} framed the system as raising visibility across teams.
Finally, participants frequently cited the value of \textcolor{rq1Color}{Multi-Signal Correlation} (4/7). By combining evidence across signals such as failed MFA, mailbox rules, and payroll changes, the system provided higher confidence. As \hyperref[M4]{M4} explained, \emph{``If we've got an outlook rule change and we see a change in direct deposit, all three of those things combined put them in a group, bam, here you go''}. Likewise, \hyperref[M2]{M2} praised the \emph{``logic looking at these different sources of intelligence''} as central to the system’s utility.

\textit{\textbf{Impact on Security Awareness:}}
Participants reported a spectrum of effects on their security awareness. The most prominent theme was \textcolor{rq1Color}{Heightened Awareness} (5/7). For example, \hyperref[M4]{M4} said the system \emph{``probably shines brighter… making us aware that somebody has been affected by that kind of attack''}, adding that they are now \emph{``more alert to account breaches''} and check inbox-rule alerts immediately.  \hyperref[M5]{M5} also noted the incident exposed a \emph{``clever man-in-the-middle''} path they had not previously considered, underscoring how the system surfaced novel attack vectors. 
At the same time, participants had \textcolor{rq1Color}{Unchanged Awareness} (2/7). Despite noting increased familiarity with policy enforcement details, \hyperref[M2]{M2} stated: \emph{``In terms of being more aware of security, I’d probably say I have the same awareness''}. \hyperref[M6]{M6} echoed this by saying \emph{``I don't think it's affected my security awareness...[from] the security side of the House''}.

A related theme was the increase in \textcolor{rq1Color}{Organizational \& Team-Level Awareness} (3/7). \hyperref[M1]{M1} framed the key benefit as cultural, explaining \emph{``the fact that people are thinking about these kinds of things… raises overall security awareness at upper levels''}. They also noted the system was useful for on-boarding new hires. \hyperref[M3]{M3} reinforced that collective awareness emerged from collaboration, stating, \emph{``there's not necessarily an easy route to the security that we need, but if we get the right people together, we can come up with a process that works''}. 
Participants also pointed to \textcolor{rq1Color}{Technical \& Process-Level Gains} (3/7). \hyperref[M2]{M2} described increased awareness of \emph{``low-level implementation details that bring a security policy about''}, while \hyperref[M3]{M3} said they better understood \emph{``how much effort it is to get an alert like this in place''}. \hyperref[M6]{M6} highlighted process improvement rather than personal awareness: \emph{``Help us… limit the amount of time from compromise to detecting that compromise''}.
However, one participant (\hyperref[M5]{M5}) cautioned, \emph{``This is an ongoing arms race… there is no secure software… someone will eventually hack it''}, suggesting that while awareness and processes improved, they saw enduring limits to defensive measures.

\subsection{\hyperref[rq:challenges]{RQ3}: Integration Enhancements}

\textit{\textbf{Improvements for Agile Integration:}}
When asked how the log-based security analytics practice could be better integrated into Agile workflows, participants identified needs spanning ownership, system tuning, automation, usability, and process culture. One theme was \textcolor{rq1Color}{Ownership, Governance, and Trust} (2/7). \hyperref[M1]{M1} stressed that the capability \emph{``needs a service owner, it needs an evangelist or advocate, it needs designated Subject Matter Experts (SMEs)''}, and should be recognized as a \emph{``critical enterprise-wide service''}. Others emphasized that trust in governance structures is necessary before enabling more automated actions.
Participants also called for \textcolor{rq1Color}{System Tuning \& Enriched Multi-Signal Logic} (3/7), including formal review loops for positives and false positives, , weighted scoring models, and broader detection indicators.
Participants also pointed to \textcolor{rq1Color}{Automation \& Advanced Capabilities} (2/7), \hyperref[M5]{M5} envisioned \emph{``AI-readable rules or an AI engine developing/altering them… accounts locked down in real time''}, while \hyperref[M6]{M6} supported expanding automated coverage to reduce reliance on manual oversight.  
Practical concerns around \textcolor{rq1Color}{Integration Guidance \& Usability} (2/7) also surfaced. \hyperref[M3]{M3} asked whether \emph{``templates or example playbooks''} existed to help teams adopt the system. They also suggested expanded alerts (e.g., hosts that stop sending logs) and monitoring for pipeline uptime.  
Finally, some participants commented on \textcolor{rq1Color}{Cultural \& process factors} (2/7). For \hyperref[M2]{M2}, integration was seamless since \emph{``hard for me to think in any way other than using an agile methodology''}. 


\textit{\textbf{Support and Resources Needed for Effective Utilization:}}
When asked what resources were necessary for sustained effectiveness, the dominant concern was \textcolor{rq1Color}{Ownership, Governance, and Sustainability} (4/7). 
Participants stressed the need for clearly assigned responsibility, cross-team authority, and dedicated engineering time to prevent the system from degrading over time. 
 \hyperref[M2]{M2} stated: \emph{``Ongoing maintenance is the big problem… software is prone to rust… needs to be maintained or it will stop working''}. 
A second major concern was \textcolor{rq1Color}{Infrastructure \& Hosting Robustness} (2/7). \hyperref[M1]{M1} called for enterprise-grade support such as backup machines, cluster-level load balancing, and failover, noting that \emph{``we can’t make it to a critical service until we get those things''}. \hyperref[M5]{M5} flagged risks in the current setup: the host server was shared by multiple developers, subject to reboots, and lacking strict access controls---conditions they deemed unsuitable for production.  
The need for \textcolor{rq1Color}{Training, Documentation, \& Knowledge Building} (2/7) was also raised. \hyperref[M3]{M3} recommended \emph{``a little introductory Zoom explaining what the detection system does, and then some documents on how to implement it''}. Similarly \hyperref[M4]{M4} requested more Splunk related training.
Participants pointed to \textcolor{rq1Color}{Expansion, Adaptability, \& Resource Needs} (3/7). \hyperref[M2]{M2} advocated for generalizing the framework beyond fraud detection to \emph{``flag risky behaviors… like spam scoring based on criteria''}, while reiterating the resource requirement for broadening scope. \hyperref[M6]{M6} agreed, calling additional engineering time the main constraint on future expansion.  
Finally, \textcolor{rq1Color}{User Feedback \& Iterative Improvement} (1/7) was identified by \hyperref[M7]{M7}, who recommended collecting suggestions from end users after deployment and using these insights to surface flaws and refine the system.  



\textit{\textbf{Redesign Considerations:}}
Most participants expressed \textcolor{rq1Color}{General Satisfaction with the Current Process} (4/7), describing the rollout as deliberate and well-coordinated. As \hyperref[M3]{M3} noted, \emph{``the process was good, slow, but deliberate, with planning, pre-prod, testing, and live runs''}. 
Rather than structural redesign, participants suggested incremental refinements.
Participants recommended refinements to \textcolor{rq1Color}{Rules, Signals, \& Detection Criteria} (3/7), such as revisiting mandatory signal combinations, expanding geographic anomaly detection, and evolving toward a broader ``System 2.0.''
Some highlighted \textcolor{rq1Color}{Governance \& Communication Improvements} (2/7). \hyperref[M2]{M2} expressed discomfort with current notification setup using Google Groups\footnote{\url{https://groups.google.com/}}, calling it \emph{``a phishing risk''}, and recommended supporting multi-recipient email functionality. One participant emphasized \textcolor{rq1Color}{Integration with Standard Processes \& Automation} (1/7). \hyperref[M5]{M5} advocated updating the system so it \emph{``creates an incident in ServiceNow''}, arguing that bringing alerts into the normal incident-response workflow would increase awareness and sustainability: \emph{``If it’s quiet for a while… people will forget… integrating into standard processing… is a good overarching strategy''}.
And another pointed to \textcolor{rq1Color}{Minor Implementation Tweaks} (1/7). \hyperref[M2]{M2} agreed with the overall architecture but noted that they would adjust details such as email handling.

\textit{\textbf{Future Features for System 2.0:}}
When asked about recommendations for a ``System 2.0,'' several highlighted the importance of \textcolor{rq1Color}{Maturity, Robustness, and Performance} (1/7). \hyperref[M1]{M1} stressed that before adding new features, the system must be load balanced, clustered, and capable of recovering without data loss. They also urged a review of pipeline latency to avoid missing delayed events. A strong theme was \textcolor{rq1Color}{Ownership, Governance, and Support} (3/7). \hyperref[M1]{M1} argued for \emph{``official support… a service owner… a list of skills and SMEs,}” while both \hyperref[M5]{M5} and \hyperref[M6]{M6} noted that for the system to become a supported service, \emph{``there have to be people in charge… probably (another division) should own it''}. Participants also called for \textcolor{rq1Color}{Automation, Integration, and Notification} (3/7). \hyperref[M1]{M1} suggested linking alerts directly to Service-Now~\footnote{\url{https://www.servicenow.com/}} and paging systems. \hyperref[M2]{M2} recommended adopting a more configuration-driven and infrastructure-as-code design, including message templating and scripts to insert alerts via API. \hyperref[M4]{M4} urged moving beyond email as the primary notification channel, proposing \emph{``Slack integration is fine… but everybody does have a Teams licence''} and advocating for a lightweight web application to track flagged users. Another major theme was \textcolor{rq1Color}{Expanded Signals and Monitoring Scope} (4/7). \hyperref[M4]{M4} recommended incorporating \emph{``risky sign-in''} signals, while \hyperref[M5]{M5} and \hyperref[M6]{M6} suggested broadening coverage beyond direct deposit to include other sensitive changes such as tax forms or mailing addresses. \hyperref[M3]{M3} anticipated that success with one use case would prompt requests to monitor additional processes. Participants also emphasized \textcolor{rq1Color}{Usability \& Feedback} (4/7). \hyperref[M2]{M2} noted that maintainability would be improved through clear documentation, identifying all moving parts, and centralizing configuration. \hyperref[M3]{M3} described the initiative as \emph{``a good model for a cross-unit team''}, and \hyperref[M7]{M7} stressed the importance of gathering production feedback before iterating further. Finally, two participants underscored the need for \textcolor{rq1Color}{Data Discovery \& Cataloging} (2/7). Both \hyperref[M5]{M5} and \hyperref[M6]{M6} recommended building a catalog of available Splunk data to guide monitoring.  


\section{Discussion} \label{sec:discussion}


\emph{\textbf{Sustainability and Ownership of Security Practices:}}
\label{subsec:sustainability-ownership}
Our results show uniformly high willingness to adopt and continue using the log-based fraud detection security practice (\hyperref[rq:adoption]{RQ1}), but that willingness is \emph{conditional} on clear ownership, elevating governance, reliable hosting, and sustained care. Participants repeatedly tied long-term viability to a designated service owner and SMEs, production-grade infrastructure with redundancy and monitoring, and a funded maintenance path. In the absence of these, the system is fragile, vulnerable to log-schema drift and organizational reassignments. This finding echoes prior observations that security practices persist when responsibility is explicit and embedded in roles and rituals rather than reliant on informal heroics~\cite{rajapakse2021empirical, lopez2019hopefully, poller2017can}. Interviewees emphasized that sustaining the system requires a shift from ``project'' to ``service'': once positioned as a critical enterprise service, detection capabilities must be audited, monitored, and budgeted with uptime Service Level Objectives (SLOs), change controls, and post-incident reviews to tune rules and reduce false positives.
Equally important is governance and trust. Participants linked any future move toward automation---such as locking accounts or triggering incident response---to organizational trust structures, emphasizing that governance must clearly define who is authorized to act on alerts, under what evidence thresholds, and with what rollback procedures. This reframes the technical pipeline into an enterprise process that balances automation with accountability. Without it, automation risks creating new frictions and undermining adoption. 
Regular rituals, such as monthly reviews to evaluate alerts and false positives, quarterly health checks for coverage and dependencies, and short onboarding sessions for new team members, embed the practice into the organizational cadence and prevent drift. Prior work similarly stresses that expert involvement and recurring feedback loops support lasting adoption~\cite{poller2017can, rajapakse2021empirical}.


\emph{\textbf{Balancing Integration with Agile Workflow and Velocity:}}
\label{subsec:agile-velocity}
A recurring theme in our findings was the delicate balance between embedding fraud detection capabilities into Agile workflows and preserving team velocity (\hyperref[rq:challenges]{RQ2}). On one hand, most participants reported minimal disruption to their daily development activities, noting that the system was ``baked into routine'' and caused little to no change in how they delivered features. The light operational footprint was largely due to 
the fact that much of the technical implementation was handled by a dedicated engineer, enabling other members to continue working without interruption, cautioning that once this person rotated off, the team would ``lose capability.'' On the other hand, participants acknowledged that sustainability was contingent on ongoing resource allocation. 
This illustrates a broader pattern where security integrations are often seen as ``free'' only when resourcing is temporarily subsidized, creating hidden fragility when viewed across longer time horizons.

Despite minimal day-to-day disruption, coordination and scheduling overhead emerged as challenges. Participants described weekly meetings as both a necessary and tolerable cost of adoption, while others characterized them as inefficient and distractions. 
This ``price of doing business'' in cross-unit initiatives was accepted by some participants as inevitable, but others expressed frustration with bottlenecks, decision rollover into subsequent weeks, and the need for stronger project management. These challenges underscore that integrating new security practices is not solely a technical endeavor, but one that requires careful orchestration of people and schedules.

Three participants reported that the system freed time rather than consumed it, especially once the initial plumbing was in place. By automating fraud detection through Splunk alerts, Grouper integration, and email notifications, the capability reduced reliance on manual monitoring and allowed staff to shift attention toward higher-value tasks. This efficiency gain highlights the dual nature of integration: while upfront coordination and resource trade-offs can strain Agile velocity, automation can create offsetting benefits by reclaiming analyst and developer capacity. The long-term balance depends on whether organizations institutionalize the system into normal incident-response workflows and allocate durable resources for tuning and governance. If positioned as a one-off project, the added coordination and resourcing overhead may erode velocity over time; but if embedded as a sustained service with automation and clear ownership, security practices can coexist with Agile priorities without undermining delivery speed. These findings reinforce prior evidence that security practices thrive when seamlessly aligned with existing workflows and supported by predictable resources rather than episodic, ad hoc efforts~\cite{rindell2018aligning, rajapakse2021empirical}.

\emph{\textbf{Security Benefits vs. Fragility \& False Positives:}}
\label{subsec:benefits-fragility}
Participants generally agreed that the fraud detection system improved the security posture of the organization, though they framed the benefit as conditional and still maturing (\hyperref[rq:challenges]{RQ2}). Several noted that the ability to correlate signals provided a meaningful step forward in detection capability. The system accelerated awareness of suspicious activity, reduced the time from compromise to detection, and created a formal alerting pipeline that previously did not exist. Beyond direct detection, participants valued the visibility and collaboration the system fostered across teams.
 At the same time, some emphasized that the system still relied heavily on human judgment to interpret and act on notifications, warning that without continued refinement, stakeholders could ignore or misinterpret alerts. 
This tension between benefit and fragility underscores a broader challenge in embedding security analytics into Agile workflows. 
To sustain positive impact, organizations must invest in iterative tuning, reduce reliance on manual interpretation, and build processes to validate and calibrate detections over time. 
False positives should not be dismissed as mere noise but tracked systematically to refine rule sets and improve precision. Likewise, system fragility should be managed with practices such as synthetic testing.
This finding resonates with prior studies showing that tool adoption alone is insufficient unless paired with continuous improvement and developer trust~\cite{smith2020can, rajapakse2021empirical}.

\emph{\textbf{Cultural and Awareness Shifts in Security:}}
\label{subsec:culture-awareness}
Beyond technical outcomes, our findings highlight the cultural and awareness shifts that accompany the integration of fraud detection into Agile workflows (\hyperref[rq:challenges]{RQ2}). Many participants described a heightened vigilance after deployment, noting that the system makes the breaches more visible and prompts them to check alerts more quickly. Others reported that the alerts deepened their understanding of potential attack vectors
which they had not previously considered. This shift reflects a broader security benefit: even when detections are infrequent, the presence of automated monitoring creates a persistent reminder that account compromise is an active risk. Several participants extended this impact beyond the individual level, emphasizing that the system elevated organizational awareness.

However, some participants explicitly stated that their personal awareness had not changed, often because they already operated from a security-focused role or felt insulated from the details of fraud detection. For these individuals, the system was valuable primarily as a process improvement rather than a personal learning opportunity. Others expressed ambivalence, acknowledging greater organizational awareness but describing little difference in their own security posture. This uneven distribution of cultural change is consistent with prior work showing that developers often view security as secondary to their core responsibilities~\cite{arpit-use-case}, while only a subset internalize new practices into day-to-day routines~\cite{lopez2019hopefully, green2016developers}. The result is a culture where awareness is raised collectively but does not uniformly translate into individual behavior change.

A final theme in participants’ reflections was the recognition that awareness alone is insufficient in the face of a constantly evolving threat landscape. 
This leads to the acknowledgment that,
while integrating fraud detection improved visibility and collaboration, it cannot eliminate risk. The implication is that organizations must treat cultural awareness as necessary but not sufficient, embedding reinforcement mechanisms such as regular security reviews, training sessions, or gamified exercises can sustain engagement over time~\cite{weir2023incorporating, triantafyllou2022gamification}. In this way, cultural change complements the technical and governance investments required for long-term resilience.

\emph{\textbf{Pathways to System 2.0: Automation, Expansion, and Usability:}}
\label{subsec:system2}
Looking forward, participants envisioned a ``System 2.0'' that would mature the current implementation into a robust enterprise service (\hyperref[rq:improvement]{RQ3}). They emphasized that before adding new capabilities, the system must reach production-grade stability with redundancy, clustering, and latency controls. 
Alongside robustness, ownership and governance emerged as central: participants argued that the system requires an official service owner, supported by SMEs across identity, operations, and finance, and anchored within an organizational unit capable of providing ongoing funding and support. These findings align with prior evidence that sustained adoption depends not only on technical design but also on durable service models and institutional buy-in~\cite{rajapakse2021empirical, poller2017can}.

Participants also proposed new directions for automation and integration. Some envisioned linking alerts directly to incident-management systems such as ServiceNow, arguing that embedding alerts into established workflows would both increase visibility and reduce the risk of alerts being ignored. 
Moving beyond email, participants advocated for richer notification channels, such as Teams or Slack, and lightweight dashboards that track flagged accounts in real time. These recommendations reflect a desire to reduce reliance on brittle manual oversight and integrate fraud detection into the ``muscle memory'' of existing Agile and incident-response processes, consistent with calls in the literature to align security practices with developer workflows~\cite{rindell2018aligning, smith2020can}.

Finally, expansion of scope and usability improvements were prominent themes. Several participants urged broadening the system to monitor additional sensitive processes, such as tax form changes or address updates, while others called for improved usability through better documentation, catalogues of available data, and centralized configuration management. 
Taken together, these ideas illustrate a forward trajectory in which log-based fraud detection evolves from a narrow proof-of-concept to a generalized framework for security analytics. Achieving this trajectory will require equal attention to resilience, governance, automation, and usability, underscoring that technical scaling and organizational adoption must proceed hand in hand.

\section{Related Work}
\label{section-related}

Prior proposes lightweight, incremental approaches to bring security into fast-moving development cycles, including risk- and policy-driven techniques for Agile settings~\cite{ge2006agile}, secure software lifecycles adapted to rapid delivery~\cite{williams2019secure}, and approaches for continuous security testing in CI/CD pipelines~\cite{rangnau2020continuous}. The broader movement of DevSecOps emphasizes embedding security as a first-class concern in development and operations and provides organizational and tooling patterns to support that goal~\cite{koskinen2019devsecops}. Foundational studies on Agile web development and mapping security processes to Agile frameworks (e.g., AWDWF, secure-Scrum mappings) demonstrate a range of proposals for aligning security activities with iterative workflows~\cite{hu2008agile,maier2017towards}.
Despite this progress, multiple empirical and conceptual studies report persistent frictions when bringing security into Agile and CI/CD practices. Authors have identified process misalignment, documentation and compliance overhead, and difficulties in triaging and actioning security findings as recurring obstacles~\cite{zaydi2024agile,riisom2018software,bajpai2022secure}. Surveys and practitioner studies likewise report skepticism about the compatibility of traditional, sequential security engineering with iterative development, and highlight organizational and tooling barriers that hinder day-to-day adoption of security practices~\cite{rindell2017busting,bartsch2011practitioners,ur2016software}.
Prior qualitative work (e.g.,~\cite{arpit}) emphasizes developer perceptions of tool friction and the cognitive load introduced by security alerts, motivating research that explores how security tools can be designed and integrated with minimal disruption to developer velocity.

Beyond integration aspects, prior studies investigate the technical underpinnings of fraud detection. Luell et al.~\cite{luell2010employee} propose a unifying framework for analytical fraud detection and reflected on the challenges of transferring research prototypes into production at a financial institution~\cite{luell2010employee}. Likewise, recent studies examining Splunk document how Security Information and Event Management (SIEM) platforms provide correlation searches, dashboards, adaptive response, and machine-learning driven anomaly detection---capabilities that closely match the Splunk signals and alerting mechanisms we used in our implementation and that inform practical deployment choices~\cite{shelke2025exploring}. However, empirical studies that examine how teams integrate log-driven detection into Agile workflows are comparatively scarce. Our work addresses this gap by reporting on the experiences of Splunk-centered, log-based fraud detection integration into a Kanban team.

\section{Threats to Validity \& Future Work}
Our study on the integration of log-based fraud detection in Agile workflows has some threats to validity. \textbf{Internal validity} concerns arise as our findings rely on semi-structured interviews and qualitative coding, which are susceptible to recall and social-desirability bias. The first author’s dual role (researcher and embedded engineer) may have also influenced responses. To mitigate these, we ensured participants their responses would be confidential and two independent researchers---one internal and one external to the development team. 
Future studies can incorporate other objective measures, such as continuous monitoring, to provide more rigorous evaluations of fraud detection integration in real operational conditions. 
\textbf{External validity} concerns arise as our results come from a single organization and a Kanban-oriented team responding to a specific incident; they may not generalize to other Agile modalities, industries, or log-management stacks. Replication across multiple teams and contexts is needed to assess broader applicability.
Future research should investigate log-based fraud detection or other security practice adoptions in Scrum, hybrid Agile–DevOps pipelines, and environments with different cultural or regulatory pressures. Comparative case studies would reveal how workflow structures and team norms shape sustainability.
Finally, \textbf{Construct and conclusion validity} concerns arise as we measure adoption, impact, and process improvement via practitioner perceptions rather than direct outcomes. Effectiveness remains conditional on false-positive control and human oversight, and qualitative coding introduces interpretation risk. Because the system is in its early stages, conclusions about sustainability and scale should be cautious; 
future work will incorporate operational metrics (e.g., false-positive rates, detection lead time) and longitudinal evaluation of sustained adoption. Combining practitioner interviews with quantitative logs would provide a fuller picture of security benefit and operational cost. In parallel, future iterations of the system should explore integration with enterprise incident-response workflows (e.g., ServiceNow, Splunk dashboards) and assess how governance models, automation, and usability improvements affect both adoption and resilience.

\section{Conclusion}
This paper presents an experience report on the integration of log-based fraud detection in an Agile (Kanban) development workflow following a real security incident. Our findings show that practitioner willingness to adopt and continue using the capability was uniformly high, rooted in the system’s utility for correlating suspicious events and raising organizational awareness. At the same time, most participants reported minimal disruption to day-to-day development activities, perceiving the system to deliver clear security benefits such as accelerated detection, cross-team collaboration, and cultural awareness. Yet, its effectiveness remained tempered by risks of false positives, schema drift, and reliance on manual oversight. 
These results highlight both the promise and the fragility of embedding log-based fraud detection into Agile workflows. By providing lessons learned from this experience report, we aim to provide actionable insights for practitioners seeking to integrate log-based fraud detection into modern software development and increase understanding of how this security practice can be embedded into workflows.

\bibliographystyle{ACM-Reference-Format}
\bibliography{sample-base}
\end{document}